\documentstyle[epsf]{article}
\begin{document}
\title{
Renormalization Group Relations and searching for Abelian
$Z^\prime$ Boson in the four-fermionic processes}
\author{
A. V. Gulov
\thanks{Email: gulov@ff.dsu.dp.ua}
and V. V. Skalozub
\thanks{Email: skalozub@ff.dsu.dp.ua}
\\ \it
Dniepropetrovsk State University, \\ \it Dniepropetrovsk, 49625
Ukraine}
\date{17 January 2000}
\maketitle
\begin{abstract}
The method of searching for signals of heavy virtual particles is
developed. It is based on a renormalization group equation for the
low energy effective Lagrangian and the decoupling theorem. As an
application, the model independent search for Abelian $Z^\prime$
boson in four-fermion processes is analyzed. The basic one-loop
renormalization group relation for the parameters of the effective
Lagrangian is derived which gives possibility to reduce the
problem to the scattering of the Standard Model particles in the
``external field'' substituting heavy virtual $Z^\prime$ state.
From the set of derived relations it is determined that the
absolute value of the $Z^\prime$ coupling to the axial currents
has to be the same for all fermions and is strongly correlated to
the $Z^\prime$ coupling to the scalar fields. The corresponding
dependences between the parameters of the effective Lagrangians
are derived.
\end{abstract}
%%%%%%%%%%%%%%%%%%%%%%%%%%%%%%%%%%%%%%%%%%%%%%%%%%%%%%%%%%%%%%%%%%%%%
%
%  Section 1
%
%%%%%%%%%%%%%%%%%%%%%%%%%%%%%%%%%%%%%%%%%%%%%%%%%%%%%%%%%%%%%%%%%%%%%
\section{Introduction}\label{sec:intr}

An important problem of nowadays high energy physics is searching
for deviations from the Standard Model (SM) of elementary
particles. At energies of the present day accelerators, the
deviations may appear due to virtual states of new heavy particles
entering the extended models and having the masses $\Lambda_i$
much greater than the $W$-boson mass: $\Lambda_i\gg m_W$. One of
approaches to describe contributions of these heavy states is
construction of the effective Lagrangians (EL) \cite{wudka}.

It is generally believed that the EL result from the decoupling of
contributions from virtual states of heavy particles to the
amplitudes of scattering processes featuring SM particles. Below,
we employ a traditional definition of the EL as the sum of
scattering amplitudes of SM particles at external momenta $p_{\rm
ext}\ll\Lambda_i$. In the case of this definition, the
contributions of virtual heavy particles to the amplitudes
decouple in the form of local operators involving SM fields, their
derivatives, and masses. These operators, ${\cal O}_i$, by
construction respect gauge symmetries of the SM, have the
dimension $d>4$, and are suppressed by the appropriate powers of
heavy masses:
\begin{equation}\label{intr:1}
 {\cal L}_{\rm eff}=
 {\cal L}^{SM}_{\rm eff}
 +\sum\limits_i\frac{\alpha_i{\cal O}_i}{\Lambda^2}
 +O\left(\frac{m^4_W}{\Lambda^4}\right).
\end{equation}
The dimensionless parameters $\alpha_i$ characterize physics
beyond the SM. They are model-dependent and expressed in terms of
dimensionless coupling constants.

In this way, one can write down the EL of type (\ref{intr:1}) for
any model supposed as the basis low-energy theory. For instance,
the minimal super-symmetric SM can be considered. In this case the
content of operators ${\cal O}_i$ must be substituted
\cite{mssm-el}. In what follows, we choose the minimal SM as the
basis low-energy theory.

All the types of operators ${\cal O}_i$ that may arise in specific
renormalizable gauge theories, containing the SM as a subgroup,
are investigated in Refs. \cite{arzt}. Then, the EL is constructed
as the sum of the type (\ref{intr:1}) with the coefficients
$\alpha_i$ treated as independent parameters to be determined from
experimental data. This meaning of $\alpha_i$ is peculiar to the
approaches used in Refs. \cite{wudka}. In contrast, the parameters
$\alpha_i$ entering Eq. (\ref{intr:1}) and defined via the
scattering amplitudes are not independent in general. For such a
definition, a set of relations between $\alpha_i$ has been derived
and called the renormalization group (RG) relations \cite{hhiggs}.
It is a consequence of the RG equations for $S$-matrix elements
and the decoupling theorem \cite{decoupling}. In what follows, we
extend the analysis to the case of searching for heavy $Z^\prime$
boson in four-fermionic processes. We suppose that the effective
(or basis) low-energy theory (the minimal SM in our case) is
specified. That means we choose the appropriate set of operators
${\cal O}_i$ in Eq. (\ref{intr:1}). We discriminate between the
Abelian and the non-Abelian $Z^\prime$ bosons, assuming the
different types of their interactions with light particles.

The low-energy model is the remnant of some unknown high-energy
theory predicting the Abelian $Z^\prime$ boson. The constraints on
the $Z^\prime$ parameters can be derived with or without knowledge
of the underlying theory. In this regard, we will call them the
model dependent and the model independent ones \cite{leike}. The
RG relations being the consequence of renormalizability are the
model independent constraints.

Let us summarize the main steps in deriving the RG relations
\cite{hhiggs}. First, as the decoupling theorem guarantees, the
operators originating from interactions of the renormalizable
types have to be preserved in the EL in leading order in
$1/\Lambda$. Since the underlying theory is not specified, the
corresponding $\alpha_i$ must be considered as arbitrary
parameters. Hence, one is able at least to relate different
parameters describing different scattering processes. Second, a
heavy virtual state can be treated as an external field scattering
SM light particles. The transition amplitude is splitted in two
scattering vertices with arbitrary vertex coefficients. The
renormalizability of the theory can be used in a way to
incorporate information about the $\beta$ and $\gamma$ functions
entering the RG operator at low energies in the framework of the
external field problem. The key point of the renormalizability of
the vertices exploited is that the algebraic structure of the
divergent part of a scattering amplitude calculated in higher
orders of perturbation theory coincides with that of the
tree-level vertex. Hence, the relation between the vertex
coefficient and the $\beta$ and $\gamma$ functions follows. If the
underlying theory is specified, the relation is just an identity.
But if the theory is not assumed and the number of $\beta$ and
$\gamma$ functions is less than the number of RG relations, a
non-trivial system of algebraic equations for vertex coefficients
and, hence, for parameters $\alpha_i$ in Eq. (\ref{intr:1}) can be
obtained.

Although the underlying theory predicting the Abelian $Z^\prime$
boson is not specified, it is possible to parameterize the
$Z^\prime$ interactions with the SM particles assuming the
effective gauge group to be ${\rm SU}(2)_L\times{\rm
U}(1)_Y\times\tilde{\rm U}(1)$ as it is often discussed in
literature \cite{sirlin,cvetic}. This parameterization is natural,
in regard of the decoupling theorem, when the leading order
operators $\sim 1/\Lambda^2$ are investigated. In general, there
are not reasons to assume a priori that the complete Lagrangian of
the model is the $\tilde{\rm U}(1)$ invariant. However, for our
approach the low-energy parameterization is important. As it will
be seen, the derived RG relations distinguish the Abelian
$Z^\prime$ and the most general parameterizations of the
$Z^\prime$ interactions with the SM fields. Moreover, in the
latter case the subgroup of $Z^\prime$ gauge group is fixed (due
to the renormalizability).

In what follows, we will formulate the low-energy RG relations for
four-fermion amplitudes in order $\sim 1/\Lambda^2$. As it will be
shown, the relations hold only in the case if the absolute value
of the $Z^\prime$ coupling to the axial-vector currents is the
same for all the SM fermions. Hence, some of the parameters
$\alpha_i$ are not independent. Moreover, being a consequence of
the renormalizability, the RG relations ensure that the Yukawa
terms of the SM Lagrangian respect the $\tilde{\rm U}(1)$ gauge
symmetry associated with the $Z^\prime$ boson.

The content is as follows. In Sec. \ref{sec:model} the model
investigated is described. In Sec. \ref{sec:RGrelations} the
general RG relation for a $S$-matrix element at low energies is
derived. In Sec. \ref{sec:RG4ferm} it is applied to four-fermion
scattering processes to obtain correlations between the vertex
parameters describing new physics at low energies. In Sec.
\ref{sec:EL} the corresponding relations for the parameters
$\alpha_i$ of the EL are derived. The results of the investigation
as well as the prospects are discussed in Sec. \ref{sec:discuss}.

%%%%%%%%%%%%%%%%%%%%%%%%%%%%%%%%%%%%%%%%%%%%%%%%%%%%%%%%%%%%%%%%%%%%%
%
%  Section 2
%
%%%%%%%%%%%%%%%%%%%%%%%%%%%%%%%%%%%%%%%%%%%%%%%%%%%%%%%%%%%%%%%%%%%%%
\section{The Model}\label{sec:model}

The model under consideration is assumed to be the low energy
remnant of some unknown underlying theory. It involves the SM and
the additional Abelian vector boson $Z^\prime$ which is supposed
to be a massive particle with the mass $\Lambda_0$ much heavier
than the $W$-boson mass. In the present paper the mechanism of
heavy mass generation is not considered since the underlying
theory describing interactions at energies $E\sim\Lambda_0$ is
unspecified. So, the Lagrangian includes the explicit mass term
for the $Z^\prime$ boson, and the mass $\Lambda_0$ is treated as a
parameter of the model.

The most general parameterization of the interactions between the
Abelian $Z^\prime$ and the SM fields can be performed on the base
of the effective gauge group ${\rm SU}(2)_L\times{\rm
U}(1)_Y\times\tilde{\rm U}(1)$ \cite{sirlin,cvetic}. Then, the
corresponding Lagrangian is
\begin{eqnarray}\label{model:1}
  {\cal L}&=& \frac{1}{2}{\left|\left(
  \partial_\mu
  -\frac{i g}{2}\sigma^a A^a_\mu
  -\frac{i g^\prime}{2}B_\mu
  -i\tilde{g}\tilde{Y}_\phi Z_{0\mu}^\prime \right)\phi\right|}^2
   \nonumber\\
  &&+\sum\limits_{f_L}\sum\limits_{i,j=1,2}
  \bar{f_L}_i \left(
   i\partial_\mu I_{ij}
   +g\frac{\sigma^a_{ij}}{2}A_\mu^a
   \right.\nonumber\\&&+\left.
   g^\prime \frac{y_L}{2}I_{ij}B_\mu
   +\tilde{g}\tilde{Y}^L_{f_i}I_{ij}Z_{0\mu}^\prime \right)
   \gamma^\mu{f_L}_j
   \nonumber\\
  &&+\sum\limits_{f_R}
    \bar{f_R} \left(
    i\partial_\mu
    +g^\prime Q_f B_\mu
    +\tilde{g}\tilde{Y}^R_f Z_{0\mu}^\prime \right) \gamma^\mu f_R,
\end{eqnarray}
where $\phi$ is the SM scalar doublet, and the summation over all
the SM left-handed doublets $f_L$ and the SM right-handed singlets
$f_R$ is understood. The charges $g, g^\prime, \tilde{g}$ and the
fields $A_\mu^a$, $B_\mu$, $Z_{0\mu}^\prime$ correspond to the
gauge groups ${\rm SU}(2)_L$, ${\rm U}(1)_Y$, $\tilde{\rm U}(1)$,
respectively, $Q_f$ is the fermion charge in the positron charge
units, $\sigma^a$ are the Pauli matrices, $I_{ij}$ is a $2\times
2$ unity matrix, and $y_L=-1$ for leptons and $y_L=1/3$ for
quarks. The values of the dimensionless constants $\tilde{Y}^L_f$,
$\tilde{Y}^R_f$, $\tilde{Y}_\phi$ depend on the particular
higher-energy theory including $Z^\prime$ boson. Here, they are to
be considered as arbitrary parameters. This parameterization, for
instance, accounts for the most general $Z^\prime$ interactions
generated in string theories \cite{cvetic}.

Up to this point the electroweak symmetry ${\rm SU}(2)_L\times
{\rm U}(1)_Y$ was supposed to be unbroken, and all the SM
particles were massless. The masses of the SM particles are
generated by the spontaneous breakdown of the symmetry ${\rm
SU}(2)_L\times{\rm U}(1)_Y\to{\rm U}(1)_{\rm em}$ originated due
to the non-zero vacuum expectation value of the scalar doublet.
However, the mass eigenstates are appeared to be shifted away from
the original fields $A_\mu^a, B_\mu, Z_{0\mu}^\prime$, since the
corresponding mass terms become non-diagonal. The physical fields
$A_\mu, Z_\mu, Z_\mu^\prime$ are obtained by the orthogonal
transformation:
\begin{equation}\label{model:2}
 \left\{
  \begin{array}{l}
  A_\mu= A^3_\mu\sin{\theta_W}+ B_\mu\cos{\theta_W},\\
  Z_\mu= Z_{0\mu}^\prime\sin{\theta_0}
  +\left(A^3_\mu\cos{\theta_W}
   -B_\mu\sin{\theta_W}\right)\cos{\theta_0},\\
  Z^\prime_\mu= Z_{0\mu}^\prime\cos{\theta_0}
  -\left(A^3_\mu\cos{\theta_W}
   -B_\mu\sin{\theta_W}\right)\sin{\theta_0},
  \end{array}
  \right.
\end{equation}
\begin{equation}\label{model:3}
 \left\{ \begin{array}{l}
  B_{\mu}=  A_{\mu}\cos{\theta_W}-\left(
   Z_{\mu}\cos{\theta_0}
   -Z^\prime_{\mu}\sin{\theta_0}\right) \sin{\theta_W} ,\\
  A^{3}_{\mu}= A_{\mu}\sin{\theta_W} +\left(
   Z_{\mu}\cos{\theta_0} -
   Z^\prime_{\mu}\sin{\theta_0}\right) \cos{\theta_W} ,  \\
  Z^\prime_{0\mu}= Z_{\mu}\sin{\theta_0} +
   Z^\prime_{\mu}\cos{\theta_0} ,
  \end{array}
  \right.
\end{equation}
where $\theta_W$ is the SM value of the Weinberg angle
($\tan\theta_W=g^\prime/g$), and $\theta_0$ denotes the mixing
angle relating physical states $Z_\mu$, $Z_\mu^\prime$ to massive
neutral components of the ${\rm SU}(2)_L\times{\rm
U}(1)_Y\times\tilde{\rm U}(1)$ gauge fields. The value of the
angle $\theta_0$ can be determined from the relation
\begin{equation}\label{model:4}
 \tan^2\theta_0=
 \frac{m^2_W/\cos^2\theta_W-m^2_Z}
 {\Lambda^2-m^2_W/\cos^2\theta_W},
\end{equation}
confirming Ref.\cite{sirlin}.

The masses of the physical fields are given by the expressions:
\begin{eqnarray}\label{model:5}
 m^2_A&=& 0,
  \nonumber \\
 m^2_Z&=&\frac{m^2_W}{\cos^2{\theta_W}}\left(
    1-\frac{4\tilde{g}^2\tilde{Y}^2_\phi}{g^2}
    \frac{m^2_W}{\Lambda^2-m^2_W/{\cos}^2{\theta_W}}\right),
  \nonumber\\
 \Lambda^2&=&\Lambda^2_0+ \left(
    \frac{m^2_W}{\cos^2{\theta_W}}-m^2_Z\right)
    +\frac{4\tilde{g}^2\tilde{Y}^2_\phi}{g^2}m^2_W.
\end{eqnarray}

As is seen, the mass of the physical $Z$-boson differs from the SM
value $m_W/\cos\theta_W$ by the small quantity of order $\sim
m_W^2/\Lambda^2$. So, the mixing angle $\theta_0$ is also small
$\theta_0\simeq \tan\theta_0\simeq \sin\theta_0\sim
m_W^2/\Lambda^2$. The difference $m^2_Z-m^2_W/\cos^2\theta_W$ is
negative and completely determined by the $Z^\prime$ coupling to
the scalar doublet. Thus, constraints on the $Z^\prime$
interaction with scalar fields can be derived by means of
experimental detecting this observable:
\begin{equation}\label{model:5a}
  \frac{\tilde{g}^2\tilde{Y}^2_\phi}{\Lambda^2}=
  \left(1 -\frac{m^2_Z{\cos^2}{\theta_W}}{m^2_W}\right)
  \frac{g^2}{4m^2_W} +O\left(\frac{m^4_W}{\Lambda^4}\right).
\end{equation}

Using Eqs. (\ref{model:3}) the Lagrangian of the model can be
expressed in terms of the physical fields. The dependence on the
mixing $\theta_0$ in Eqs. (\ref{model:3}) causes new interactions
$\sim m^2_W/\Lambda^2$ originally absent in the Lagrangian
(\ref{model:1}).

In the present paper we will deal with four-fermion amplitudes of
order $\sim \Lambda^{-2}$ produced by virtual $Z^\prime$ boson in
the $s$ channel. The corresponding tree-level diagram is shown in
Fig. \ref{fig:1}. At low energies the effective four-fermion
vertex is generated due to the heavy mass in the $Z^\prime$-boson
propagator: ${(p^2- \Lambda^2)}^{-1}\to
-\Lambda^{-2}[1+O(p^2/\Lambda^2)]$. So, expressions involving such
amplitudes can be computed neglecting the terms proportional to
$\sim m^2_W/\Lambda^2$ in the Lagrangian. Only linear in
$Z^\prime$ terms of the Lagrangian are needed, which have the
following form in the 't Hooft-Feynman gauge:
\begin{eqnarray}\label{2.6}
 {\cal L}^\prime&=&
 \tilde{g}Z^\prime_\mu
 \sum\limits_f\bar{f}\gamma^\mu
  \left(\tilde{Y}^L_f\omega_L +\tilde{Y}^R_f\omega_R\right)f
\nonumber\\
 && +\tilde{g}\tilde{Y}_\phi Z^\prime_\mu
 \left[
  H\stackrel{\leftrightarrow}{\partial_\mu}\chi_3
  +i\chi^-\stackrel{\leftrightarrow}{\partial_\mu}\chi^+
  +2eA_\mu\chi^-\chi^+
 \right. \nonumber\\
 &&-\frac{g}{2\cos\theta_W}
  \left(
  H^2+2vH-2\chi^-\chi^+\cos{2\theta_W}+\chi^2_3\right)Z_\mu
 \nonumber\\
 &&-ig\left( H+v\right)\left(
   W^+_\mu\chi^- -W^-_\mu\chi^+\right)
 \nonumber\\
 &&\left. -g\chi_3\left(
    W^+_\mu\chi^- +W^-_\mu\chi^+\right)
 \right] +O\left(\frac{m^2_W}{\Lambda^2}\right) ,
\end{eqnarray}
where $\varphi\stackrel{\leftrightarrow}{\partial_\mu}\chi\equiv
\varphi\partial_\mu\chi-\chi\partial_\mu\varphi$,
$\omega_{L,R}=(1\mp\gamma^5)/2$, $e=g\sin\theta_W$ is the positron
charge, $v$ denotes the vacuum expectation value of the scalar
doublet, $H$ is the Higgs scalar particle, and $\chi^\pm$,
$\chi_3$ are the Goldstone fields.

%%%%%%%%%%%%%%%%%%%%%%%%%%%%%%%%%%%%%%%%%%%%%%%%%%%%%%%%%%%%%%%%%%%%%
%
%  Section 3
%
%%%%%%%%%%%%%%%%%%%%%%%%%%%%%%%%%%%%%%%%%%%%%%%%%%%%%%%%%%%%%%%%%%%%%
\section{RG relations in a theory with decoupling}
\label{sec:RGrelations}

As we mentioned before, the model under consideration is supposed
to be the low-energy limit of some unspecified renormalizable
theory describing interactions at energies $E\ge\Lambda$. As is
known, due to renormalizability the $S$-matrix elements are
invariant with respect to the RG transformations which express
independence of the amplitudes of the normalization point
$\kappa$. In a theory with different mass scales the RG flow and
the decoupling of heavy loop contributions at the thresholds of
heavy masses give possibility to determine all the parameters of
scattering amplitudes. The decoupling is resulted in an important
property of the low energy amplitudes: the running of the proper
functions is regulated by loops of light particles. Therefore, the
$\beta$ and $\gamma$ functions at low energies, $E\ll\Lambda$, are
determined by the SM particles, only. The latter fact allows to
formulate a series of the RG relations for scattering amplitudes.
We will derive them for the case of heavy $Z^\prime$ boson which,
as well as other heavy particles, is decoupled.

We will use the notations $\lambda_a= g^2, g^{\prime 2}, g_s^2,
\tilde{g}^2, G_f, \lambda$ (where $g_s$, $G_f=m_f/v$ and $\lambda$
denote the QCD charge, the coupling of the Yukawa interaction
between $f$ and $\phi$, and the scalar self-coupling,
respectively) and $X= \Phi, m_i^2,\Lambda^2$ (where $\Phi$
represents all the fields of the theory, and $m_i$ are particle
masses) to refer to the charges, the fields, and the masses. Then
the RG equation reads:
\begin{eqnarray}\label{3.7}
&& \frac{d}{d\ln\kappa}{\hat S}=0, \nonumber \\ &&
\frac{d}{d\ln\kappa}\equiv \hat{\cal
D}=\frac{\partial}{\partial\ln\kappa}+
\sum\limits_a{\hat\beta}_a\frac{\partial}{\partial{\hat\lambda}_a}-
\sum\limits_X{\hat\gamma}_X\frac{\partial}{\partial\ln\hat{X}},
\end{eqnarray}
where renormalized (running) parameters $\hat{\lambda}_a$,
$\hat{X}$ are the solutions to the equations:
\begin{equation}\label{3.8}
\frac{d{\hat{\lambda}}_a}{d\ln\kappa}={\hat\beta}_a\left(
{\hat\lambda}_a, \hat X \right), \quad \frac{d\ln\hat
X}{d\ln\kappa}= -{\hat\gamma}_X\left( {\hat\lambda}_a, \hat X
\right).
\end{equation}
Hats over quantities mark the parameters of the underlying theory.
They include the contributions of loops containing both the SM
particles and the heavy ones.

Below, in carrying out calculations the dimensional regularization
and the $\overline{\rm MS}$ renormalization scheme \cite{cheng}
will be used. The $\overline{\rm MS}$ scheme is mass independent
one, so the RG operator $\hat{\cal D}$ has the same form at high,
$E\ge\Lambda$, and low, $E\ll\Lambda$, energies.

In the energy range $E\ll\Lambda$ all the heavy fields are
decoupled. So, one can introduce the running parameters
$\lambda_a$, $X$ completely defined by the low-energy effective
theory \cite{bando}:
\begin{equation}\label{3.9}
\frac{d\lambda_a}{d\ln\kappa}=\beta_a\left(\lambda_a, X \right),
\quad \frac{d\ln X}{d\ln\kappa}= -\gamma_X\left(\lambda_a, X
\right),
\end{equation}
where $\beta$ and $\gamma$ functions contain no contributions from
loops with heavy particles. These parameters can be expressed in
terms of the original ones:
\begin{eqnarray}\label{3.10}
\lambda_a&=&{\hat\lambda}_a
+a_{\lambda_a}\ln\frac{{\hat\Lambda}^2}{\kappa^2}
+b_{\lambda_a}\ln^2\frac{{\hat\Lambda}^2}{\kappa^2}+...
\nonumber\\ X&=&\hat{X}\left( 1
+a_X\ln\frac{{\hat\Lambda}^2}{\kappa^2} +b_X
\ln^2\frac{{\hat\Lambda}^2}{\kappa^2}+...\right) ,
\end{eqnarray}
where the matching between the both sets of parameters
($\lambda_a$, $X$ and  ${\hat\lambda}_a$, $\hat{X}$) is chosen to
be done at the normalization point $\kappa\sim\Lambda$:
\begin{equation}\label{3.11}
\lambda_a\mid_{\kappa=\Lambda}
 ={\hat\lambda}_a\mid_{\kappa=\Lambda},
\quad X\mid_{\kappa=\Lambda}=\hat{X}\mid_{\kappa=\Lambda}.
\end{equation}

Actually, Eq. (\ref{3.10}) is the redefinition of the parameters
of the theory. As it has been shown in Refs. \cite{bando,2sc}, all
the heavy particle loop contributions proportional to $\ln\kappa$
are completely removed from the RG equation (\ref{3.7}) by such a
redefinition. This fact is a consequence of the decoupling theorem
\cite{decoupling}.

Then, the RG equation (\ref{3.7}) for the $\hat{S}$ matrix
expressed in terms of the low-energy quantities $\lambda_a$, $X$
becomes:
\begin{equation}\label{3.12}
{\cal D}S=\left( \frac{\partial}{\partial\ln\kappa} +\sum\limits_a
\beta_a\frac{\partial}{\partial\lambda_a} -\sum\limits_X
\gamma_X\frac{\partial}{\partial\ln{X}}\right) S=0,
\end{equation}
where ${\cal D}$ and $S$ denote, respectively, the RG operator and
the $S$-matrix element of the low-energy effective theory,
calculated without the heavy particle loops.

The familiar usage of Eq. (\ref{3.12}) is to improve scattering
amplitudes calculated in perturbation theory. In contrast, in what
follows we will apply Eq. (\ref{3.12}) to obtain algebraic
relations between the parameters $\tilde{Y}^L_f$, $\tilde{Y}^R_f$,
$\tilde{Y}_\phi$, which are to be considered as arbitrary numbers,
if the underlying theory is not assumed. The reasons for that are
as follow. In the case when the underlying theory is specified
($\tilde{Y}^L_f$, $\tilde{Y}^R_f$, $\tilde{Y}_\phi$ are to be
computed) and the $\beta$ and $\gamma$ functions as well as the
$S$-matrix elements are calculated in a fixed order of
perturbation theory, Eq. (\ref{3.12}) is just the identity. In a
sense, it is a necessary condition of renormalizability. If the
underlying theory is not specified, whereas the $\beta$, $\gamma$
functions and the $S$-matrix elements are computed in a fixed
order of perturbation theory, then, as we will show for
four-fermion processes at one-loop level, the equality
(\ref{3.12}) holds only if the parameters $\tilde{Y}^L_f$,
$\tilde{Y}^R_f$, $\tilde{Y}_\phi$ satisfy some specific
correlations. The latter result in the dependences between the
appropriate parameters $\alpha_i$ in the EL (\ref{intr:1}).

Let $S^{(1)}$ and $S^{(0)}$ represent the one-loop and the tree
parts of the $S$-matrix element, respectively. Introducing the
one-loop RG operator:
\begin{equation}\label{3.13}
{\cal D}^{(1)}\equiv \sum\limits_a
\beta^{(1)}_a\frac{\partial}{\partial\lambda_a} -\sum\limits_X
\gamma^{(1)}_X\frac{\partial}{\partial\ln{X}},
\end{equation}
where $\beta$ and $\gamma$ functions are computed in one-loop
order, one can derive from Eq. (\ref{3.12}) the following
identity:
\begin{equation}\label{3.14}
\frac{\partial}{\partial\ln\kappa}S^{(1)} +{\cal
D}^{(1)}S^{(0)}=0.
\end{equation}
Since the parameters $\lambda_a$, $X$ and ${\hat\lambda}_a$,
$\hat{X}$ differ at the one-loop level, one could freely
substitute one set of them by another in Eq. (\ref{3.14}).

Equation (\ref{3.14}) is the starting point to analyze the RG
relations for amplitudes at low energies. The relation
(\ref{3.14}) ensures the leading logarithmic terms of the one-loop
$S$-matrix element to reproduce the appropriate tree-level
structure that is a simple consequence of renormalizability. When
the couplings are represented by arbitrary unknown parameters, as
it takes place in the considered model and, in general, in the
discussed EL approach, the relation (\ref{3.14}) is the one-loop
equation for the parameters which must be determined with the
computed $\beta$ and $\gamma$ functions. If due to a symmetry the
number of $\beta$ and $\gamma$ functions is less than the number
of RG relations, the non-trivial system of equations correlating
the originally independent parameters may occur. We will use Eq.
(\ref{3.14}) to derive dependences between the parameters
$\tilde{Y}^L_f$, $\tilde{Y}^R_f$, and $\tilde{Y}_\phi$.

%%%%%%%%%%%%%%%%%%%%%%%%%%%%%%%%%%%%%%%%%%%%%%%%%%%%%%%%%%%%%%%%%%%%%
%
%  Section 4
%
%%%%%%%%%%%%%%%%%%%%%%%%%%%%%%%%%%%%%%%%%%%%%%%%%%%%%%%%%%%%%%%%%%%%%
\section{RG relations for four-fermion amplitudes}\label{sec:RG4ferm}

In this section the one-loop RG relation (\ref{3.14}) is applied
to the four-fermion processes $\bar{f}_1 f_1\to
Z^{\prime\ast}\to\bar{f}_2 f_2$ containing $Z^\prime$ boson as the
virtual state. In lower order in $\Lambda^{-2}$ such amplitudes
produce the effective contact interactions of the type
current$\times$current described by the EL (\ref{intr:1})
\cite{wudka,arzt}.

\begin{figure}
\begin{center}
  \epsfxsize=0.8\textwidth
  \epsfbox[0 0 600 600]{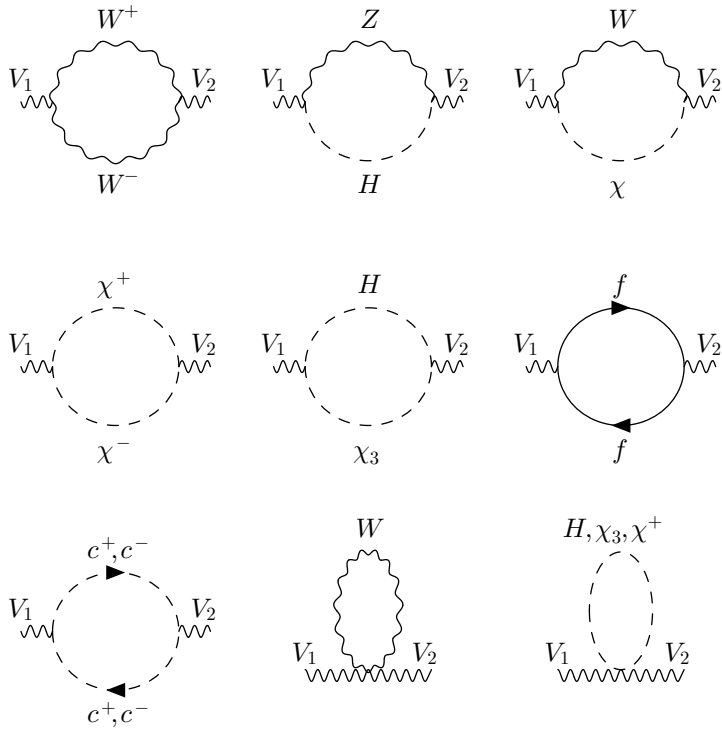}
  \caption{
   The one-loop level contributions to the polarization operator
   of neutral vector bosons ${\Pi}^{(1)}_{V_{1}V_{2}}(p^{2})$.}
\label{fig:2}
\end{center}
\end{figure}

\begin{figure}
\begin{center}
  \epsfxsize=0.8\textwidth
  \epsfbox[0 0 600 600]{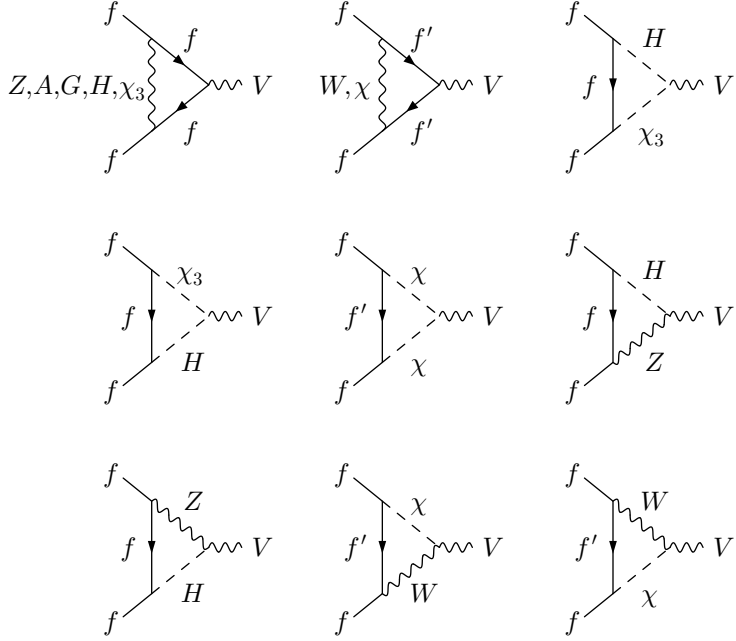}
  \caption{
   The one-loop level contributions to the vertex function
   ${\Gamma}^{(1)}_{\psi V}$.}
\label{fig:3}
\end{center}
\end{figure}

Let the renormalized fields, masses, and charges are defined as
follows:
\begin{eqnarray}\label{4.15}
 &f=Z^{-1/2}_f f^{\rm bare},&\quad
 \left(\begin{array}{c}
 A_\mu\\ Z_\mu\\ Z^\prime_\mu \end{array}\right)
 =Z^{-1/2}_V
 \left(\begin{array}{c}
 A^{\rm bare}_\mu\\ Z^{\rm bare}_\mu\\ Z^{\prime\rm bare}_\mu
 \end{array}\right),
 \nonumber\\
 &m^2_i=m^2_{i,\rm bare}-\delta m^2_i,&\quad
 \Gamma_{Vf}=\sum\limits_{V_1}
 {\left(Z^{-1}_g\right)}_{VV_1}\Gamma^{\rm bare}_{V_1 f},
\end{eqnarray}
where $\Gamma_{Vf}$ is the vertex describing interaction between
the fermion $f$ and the neutral vector boson $V$ and
${(Z^{-1/2}_V)}_{V_1 V_2}$ and ${(Z^{-1}_g)}_{V_1 V_2}$ are
$3\times 3$ matrices ($V_i=A,Z,Z^\prime$). The fermion
renormalization constant $Z^{-1/2}_f$ is the SM one. Its value is
known \cite{smvalues}. Quantities $Z^{-1/2}_V$ and $\delta m_V$
are represented at the one-loop level by the divergent terms
($\sim\varepsilon^{-1}$) of the gauge-independent part of the
polarization operator of neutral vector bosons $\Pi^{(1)}_{V_1
V_2}(p^2)$ being evaluated from the diagrams in Fig. \ref{fig:2}:
\begin{eqnarray}\label{4.16}
\left( Z^{1/2}_V\right)_{V_1 V_2}
 &=&\delta_{V_1 V_2}
  +\frac{\delta_{V_1 V_2}}{2}
   \frac{\partial}{\partial p^2}
   \Pi^{(1)div}_{V_1 V_2}\left(p^2\right)
 \nonumber\\&&
  -\frac{1-\delta_{V_1 V_2}}{m^2_{V_1}-m^2_{V_2}}
   \Pi^{(1)div}_{V_1 V_2}\left(m^2_{V_2}\right),
 \nonumber\\
\delta m^2_V&=&-\Pi^{(1)div}_{V_1 V_2}\left(m^2_V\right),
\end{eqnarray}
where $\delta_{V_1 V_2}$ is the Kronecker symbol. In order to find
the constant $Z_g$ in one-loop order one has to calculate the
divergent ($\sim\varepsilon^{-1}$) part of the vertex function
$\Gamma^{(1)}_{\psi V}$ (the diagrams in Fig. \ref{fig:3}):
\begin{equation}\label{4.17}
Z_g =Z^{-1}_f \left(Z^{-1/2}_V\right)^T
 \left(1-\Gamma^{(1)div}_{fV}/\Gamma^{(0)}_{fV}\right),
\end{equation}
where $\Gamma^{(0)}_{\psi V}$ is the vertex $\Gamma_{\psi V}$ at
the tree level.

\begin{figure}
\begin{center}
  \epsfxsize=0.5\textwidth
  \epsfbox[0 0 600 400]{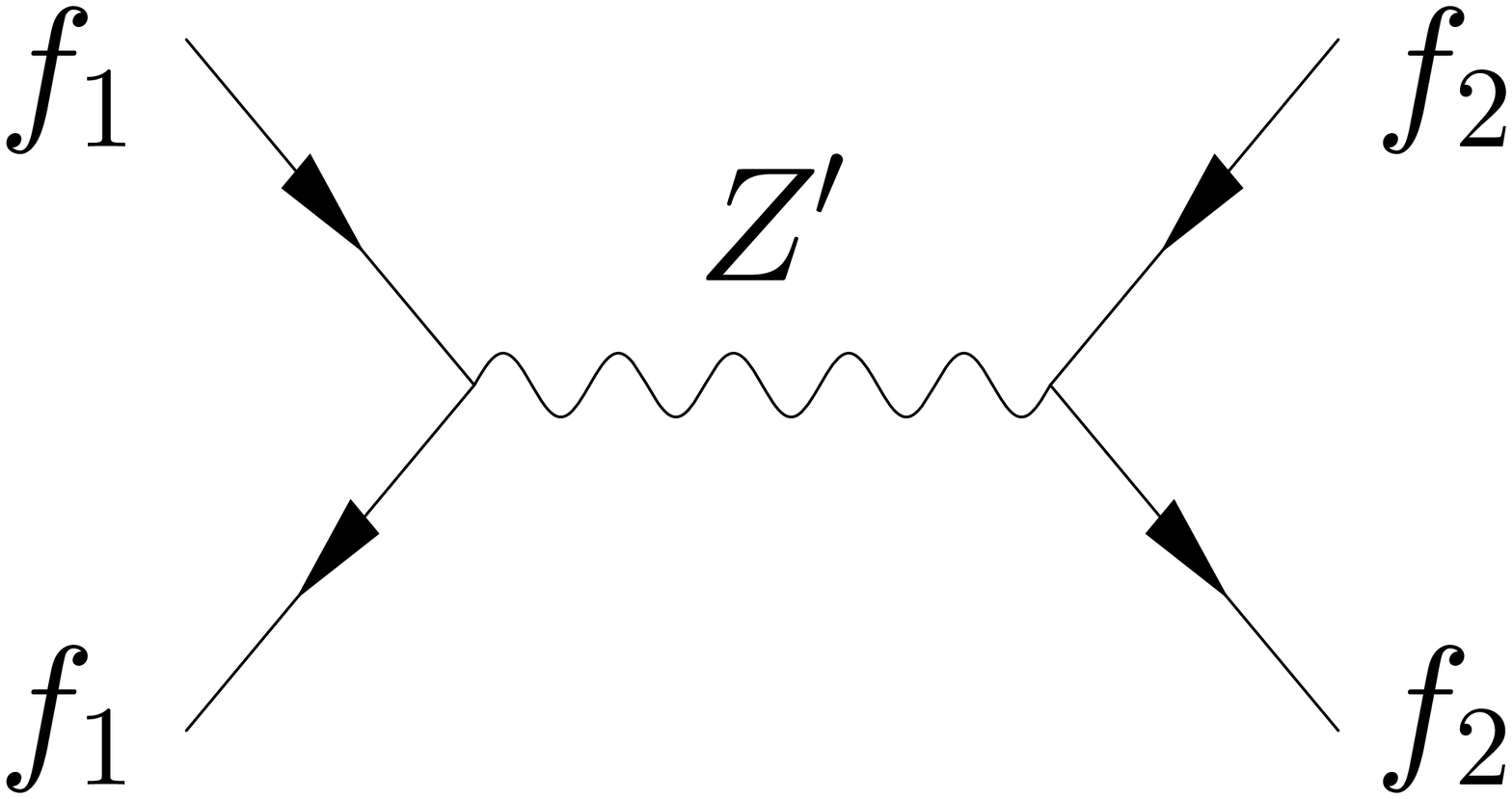}
  \caption{
   The tree-level amplitude of the process ${\bar
   f}_{1}f_{1}\to Z^{\prime \ast}\to{\bar f}_{2}f_{2}$.}
\label{fig:1}
\end{center}
\end{figure}

\begin{figure}
\begin{center}
  \epsfxsize=0.8\textwidth
  \epsfbox[0 0 600 130]{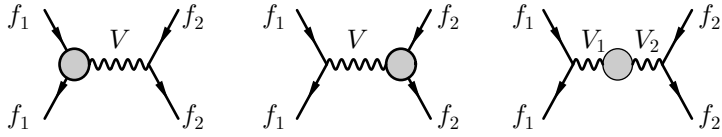}
  \caption{
   The $\ln\kappa$-dependent contributions
   to the one-loop $S$-matrix element
   describing scattering ${\bar
   f}_{1}f_{1}\to V^{\ast}\to{\bar f}_{2}f_{2}$.}
\label{fig:4}
\end{center}
\end{figure}

The $S$-matrix element expressed in terms of the renormalized
quantities is finite. It can be evaluated using the diagrams in
Figs. \ref{fig:1} and \ref{fig:4}, where shaded blobs stand for
the finite parts of the vertex function $\Gamma^{(1)fin}_{fV} =
\Gamma^{(1)}_{fV} - \Gamma^{(1)div}_{fV}$ and the polarization
operator $\Pi^{(1)fin}_{V_1 V_2} = \Pi^{(1)}_{V_1 V_2} -
\Pi^{(1)div}_{V_1 V_2}$. On the mass-shell of fermions the
gauge-dependent part of $\Pi_{V_1 V_2}$ ($\sim p_\mu p_\nu$) gives
no contribution to the physical amplitude due to transversality of
the vertex $\Gamma_{fV}$.

Because of the one-loop mixing between neutral bosons, Eq.
(\ref{3.14}) for the renormalized $S$-matrix element cannot be
reduced to relations for vertices. To derive these relations one
has to avoid the mixing. The corresponding procedure was described
in Ref. \cite{2sc}.

One has to introduce the quantities $\tilde{V}$ and
$\tilde{\Gamma}_{fV}$ instead of the fields $V=A_\mu$, $Z_\mu$,
$Z^\prime_\mu$ and vertices $\Gamma_{fV}$ according to the
following relations:
\begin{eqnarray}\label{4.18}
&&V_{1}=\sum\limits_{V_{2}}{\zeta}^{1/2}_{V_{1}V_{2}}{\tilde
V}_{2}, \quad {\Gamma}_{fV_{1}}=\sum\limits_{V_{2}}
{\zeta}^{-1/2}_{V_{2}V_{1}}{\tilde\Gamma}_{fV_{2}}, \nonumber\\
&&{\zeta}^{\pm 1/2}_{V_{1}V_{2}}={\delta}_{V_{1}V_{2}}\mp
\frac{1-{\delta}_{V_{1}V_{2}}}{m^{2}_{V_{1}}-m^{2}_{V_{2}}}
\ln\frac{\kappa}{\Lambda}\lim\limits_{\varepsilon\to
0}\varepsilon{\Pi}^{(1)}_{V_{1}V_{2}}( m^{2}_{V_{2}}) .
\end{eqnarray}

After such a redefinition the mixing between neutral bosons in Eq.
(\ref{3.14}) vanishes. In what follows, we will use the quantities
$\tilde{V}$ and $\tilde{\Gamma}_{fV}$, therefore only the diagonal
terms of the polarization operator are needed in computing of the
$S$-matrix element $S^{(1)}$. Since the difference between the two
sets of parameters is of order $\sim g^2/16\pi^2$, one can
substitute the tilded parameters by the other set in the
amplitudes in lower order.

When the diagonalization of the leading logarithmic terms of the
vector boson polarization operator is fulfilled, Eq. (\ref{3.14})
can be reduced to relations for vertices describing scattering of
fermions in the external field $\Lambda^{-1}$ substituting the
heavy $Z^\prime$ boson:
\begin{equation}\label{4.19}
 \frac{\partial{\Gamma}^{(1)}_{fV}}{\partial\ln\kappa}\frac{1}{\Lambda}
 +{\cal D}^{(1)}\left( {\Gamma}^{(0)}_{fV}\frac{1}{\Lambda}\right)
 =0.
\end{equation}

The quantities $\Gamma^{(0)}_{fV}$ and $\Gamma^{(1)}_{fV}$ in Eq.
(\ref{4.19}) have been calculated in the 't Hooft-Feynman gauge
using diagrams in Figs. \ref{fig:1}, \ref{fig:3}. The computation
gives the following relations for the left-handed and the
right-handed fermions, respectively:
\begin{eqnarray}\label{4.20}
&&\frac{1}{8\pi^2}\left[
 g^2\left(\frac{1}{4 \cos^2 \theta_W}
  +\left( Q^2_f-\left| Q_f\right| \right) \tan^2 \theta_W
  +\frac{\tilde{Y}^L_{f^\prime}}{2\tilde{Y}^L_f} \right)
 \right.\nonumber\\&&\qquad\left.
 +\frac{4}{3} g^2_{s,f}
  +G^2_f\frac{\tilde{Y}^R_f-t_f \tilde{Y}_\phi}{\tilde{Y}^L_f}
  +G^2_{f^\prime}\frac{\tilde{Y}^R_{f^\prime}
   +t_f \tilde{Y}_\phi}{\tilde{Y}^L_f}
 \right]\nonumber\\&&\qquad
 +\frac{\beta^{(1)}_{\tilde{g}}}{2\tilde{g}^2}
 +\frac{1}{2} \gamma^{(1)}_{\Lambda^2}
 -2\gamma^{(1)}_{f_L}=0,
 \nonumber\\
&&\frac{1}{8\pi^2}\left[
 g^2 Q^2_f \tan^2 \theta_W +\frac{4}{3}g^2_{s,f}
 +G^2_f\frac{\tilde{Y}^L_f+\tilde{Y}^L_{f^\prime}
  +2t_f \tilde{Y}_\phi}{\tilde{Y}^R_f}
 \right]\nonumber\\&&\qquad
 +\frac{\beta^{(1)}_{\tilde{g}}}{2\tilde{g}^2}
 +\frac{1}{2} \gamma^{(1)}_{\Lambda^2}
 -2\gamma^{(1)}_{f_R}=0,
\end{eqnarray}
where
\begin{equation}
\beta_{\tilde{g}}=\frac{d\tilde{g}^2}{d\ln\kappa}.
\end{equation}
The notation $f^\prime$ stands for the isopartner of $f$
($l^\prime=\nu_l$, ${\nu_l}^\prime=l$, ${q_d}^\prime=q_u$,
${q_u}^\prime=q_d$), $g_{s,f}$ equals to $g_s$ for quarks and zero
for leptons,
\begin{equation}
 t_f=2T^3_f=\left\{ \begin{array}{l}
  +1,f=\nu_l,u,c,t\\ -1,f=l,d,s,b
  \end{array}\right.,
\end{equation}
where $l=e,\mu,\tau$ denotes leptons, and $T^3_f$ is the third
component of the weak isospin.

The SM values of the one-loop fermion anomalous dimensions are
known \cite{smvalues}:
\begin{eqnarray}\label{4.21}
&\gamma^{(1)}_{f_L}=&\frac{1}{16\pi^2}
 \left[
  g^2\left(\frac{1}{4{\cos}^2\theta_W}
   +(Q^2_f-\left|Q_f\right|){\tan}^2\theta_W +\frac{1}{2}\right)
 \right.\nonumber\\&&\left.
  +\frac{4}{3}g^2_{s,f} +G^2_f  +G^2_{f^\prime}
 \right],\nonumber\\
&\gamma^{(1)}_{f_R}=&\frac{1}{16\pi^2}
 \left[
  g^2 Q^2_f \tan^2\theta_W +\frac{4}{3}g^2_{s,f} +2G^{2}_{f}
 \right].
\end{eqnarray}

By employing Eqs. (\ref{4.16})-(\ref{4.18}), it is easy to compute
$\beta_{\tilde{g}}$ and $\gamma_{\Lambda^2}$ in one-loop order:
\begin{eqnarray}
(\tilde{Y}^L_f)^2 \beta^{(1)}_{\tilde{g}}&=&\frac{\tilde{g}^2
\tilde{Y}^L_f}{24\pi^2}
 \left[
  2\tilde{g}^2 \tilde{Y}^L_f \Sigma +3g^2\left(\tilde{Y}^L_f
   -\tilde{Y}^L_{f^\prime}\right)
 \right.\nonumber\\&&
  +6G^2_f\left(\tilde{Y}_\phi t_f +\tilde{Y}^L_f
   -\tilde{Y}^R_f\right)
 \nonumber\\&&
  -6G^2_{f^\prime}\left(\tilde{Y}_\phi t_f
   -\tilde{Y}^L_f +\tilde{Y}^R_{f^\prime}\right)
 ]
 +O\left(\frac{m^2_W}{\Lambda^2}\right),
 \nonumber\\
(\tilde{Y}^R_f)^2 \beta^{(1)}_{\tilde{g}}
 &=&\frac{\tilde{g}^2 \tilde{Y}^R_f}{12\pi^2}
 \left[
  -3G^2_f\left(2\tilde{Y}_\phi t_f +\tilde{Y}^L_f
   +\tilde{Y}^L_{f^\prime} -2\tilde{Y}^R_f\right)
 \right.\nonumber\\&&
  +\tilde{g}^2 \tilde{Y}^R_f \Sigma
 ]
 +O\left(\frac{m^2_W}{\Lambda^2}\right),
 \label{4.22}\\
\gamma^{(1)}_{\Lambda^2}&=&-\frac{\tilde{g}^2 \Sigma}{12\pi^2},
 \label{4.23}
\end{eqnarray}
where $\Sigma=\tilde{Y}^2_\phi+ \sum\limits_f
\left((\tilde{Y}^L_f)^2+(\tilde{Y}^R_f)^2\right)$.

Since the $\beta_{\tilde{g}}$ function is the same for both Eqs.
(\ref{4.22}), the following relations can be derived:
\begin{equation}\label{4.24}
\tilde{Y}^L_{f^\prime}=\tilde{Y}^L_f,\quad
\tilde{Y}^R_f=\tilde{Y}^L_f+t_f \tilde{Y}_\phi.
\end{equation}
Hence, one can see that originally independent parameters
$\tilde{Y}^L_f$, $\tilde{Y}^R_f$, and $\tilde{Y}_\phi$ appear to
be the connected ones. Finally, one can also check that the RG
relations (\ref{4.20}) are fulfilled only if Eq. (\ref{4.24})
holds.

In a sense, the relations (\ref{4.24}) mean that the $Z^\prime$
boson couplings to the SM axial-vector currents have the universal
absolute value, if a single light scalar doublet exists. So, among
the four constants $\tilde{Y}^L_f$, $\tilde{Y}^R_f$,
$\tilde{Y}^L_{f^\prime}$, and $\tilde{Y}^R_{f^\prime}$
parameterizing the interaction between the $Z^\prime$ boson and
the ${\rm SU}(2)$ fermion isodoublet only one is really arbitrary.
The rest ones can be expressed through it and the hypercharge
$\tilde{Y}_\phi$ of the $Z^\prime$ coupling to the SM scalar
doublet. Thus, the fermion and the scalar sectors of new physics
are to be correlated.

Being derived from the condition of renormalizability of a
scattering amplitude in the external field of heavy particle, the
relations (\ref{4.24}) respect gauge invariance and have a
transparent interpretation. First of them means that that the
upper and the lower components of a left-handed doublet transform
in the same way under $\tilde{\rm U}(1)$ gauge group. If it does
not hold, ${\rm SU}(2)_L$ gauge invariance would violate. The
second of relations guarantees, that the SM Yukawa interaction
terms have to be invariant under $\tilde{\rm U}(1)$ gauge
transformations.

%%%%%%%%%%%%%%%%%%%%%%%%%%%%%%%%%%%%%%%%%%%%%%%%%%%%%%%%%%%%%%%%%%%%%
%
%  Section 5
%
%%%%%%%%%%%%%%%%%%%%%%%%%%%%%%%%%%%%%%%%%%%%%%%%%%%%%%%%%%%%%%%%%%%%%
\section{Dependences between the parameters of the effective Lagrangian}
\label{sec:EL}

The relations (\ref{4.24}) result in dependences between the
parameters of the EL (\ref{intr:1}). In Ref. \cite{arzt} all the
effective operators ${\cal O}_i$ describing deviations from the SM
and appearing at the tree level due to heavy virtual states have
been derived. The heavy $Z^\prime$ boson produces the following
effective vertices:
\begin{eqnarray}\label{5.26}
&&-\frac{\tilde{g}^2 \tilde{Y}^2_\phi}{8\Lambda^2}
  \left( \left(D^{ew}_\mu\phi\right)^\dagger\phi
   -\phi^\dagger D^{ew}_\mu \phi \right)^2
 \nonumber\\&&\qquad
 =-\frac{\tilde{g}^2 \tilde{Y}^2_\phi}{4\Lambda^2}
  \left({\cal O}_{\partial\phi}-2{\cal O}^{(3)}_\phi\right),
 \nonumber\\&&
 -\frac{i\tilde{g}^2 \tilde{Y}_\phi}{2\Lambda^2}
  \left( (D^{ew}_\mu\phi)^\dagger\phi
   -\phi^\dagger D^{ew}_\mu \phi \right)
  {\bar f}\gamma^\mu\left(\tilde{Y}^L_f \omega_L
   +\tilde{Y}^R_f \omega_R\right) f
 \nonumber\\&&\qquad
 =\frac{\tilde{g}^2 \tilde{Y}_\phi}{2\Lambda^2}
  \left({\cal O}^{(1)}_{\phi f}
   +{\cal O}_{\phi f} +\mbox{h.c.}\right),
 \nonumber\\&&
 \frac{\tilde{g}^2}{\left(1+\delta_{f_1 f_2}\right) \Lambda^2}
  {\bar f}_1 \gamma^\mu
  \left(\tilde{Y}^L_{f_1}\omega_L +\tilde{Y}^R_{f_1}\omega_R\right) f_1
 \nonumber\\&&\quad
  \times{\bar f}_2 \gamma^\mu
  \left(\tilde{Y}^L_{f_2}\omega_L +\tilde{Y}^R_{f_2}\omega_R\right) f_2
 \nonumber\\&&\qquad
 =\frac{2\tilde{g}^2}{\Lambda^2}
  \left(\frac{\tilde{Y}^L_l \tilde{Y}^L_f {\cal O}^{(1)}_{l f}
   +\tilde{Y}^R_l \tilde{Y}^R_f {\cal O}_{l f}} {1+\delta_{l f}}
   \right.
 \nonumber\\&&\qquad
  \left.
  +\frac{\tilde{Y}^L_{q_1}\tilde{Y}^L_{q_2}{\cal O}^{(1,1)}_{q_1 q_2}
   +\tilde{Y}^R_{q_1}\tilde{Y}^R_{q_2}{\cal O}^{(1)}_{q_1 q_2}}
   {1+\delta_{q_1 q_2}}\right),
\end{eqnarray}
where $l$ and $q$ stand for leptons and quarks; $D^{ew}_\mu$ is
the electroweak covariant derivative, and the definitions of
${\cal O}_i$ correspond to Ref.\cite{arzt}:
\begin{eqnarray}
&&{\cal O}_{\partial\phi}=\frac{1}{2}
  \partial_\mu\left( \phi^\dagger \phi\right)
  \partial^\mu\left( \phi^\dagger \phi\right) ,
 ~{\cal O}^{(3)}_\phi =
  {\left| \phi^\dagger D^{ew}_\mu \phi\right| }^2,
  \nonumber\\
&&{\cal O}^{(1)}_{\phi f}=i
  \left( \phi^\dagger D^{ew}_\mu \phi\right)
  \bar{f_L} \gamma^\mu f_{L},
 ~{\cal O}_{\phi f}= i
  \left( \phi^\dagger D^{ew}_\mu \phi\right)
  \bar{f_R} \gamma^\mu f_{R}, \nonumber\\
&&{\cal O}^{(1,1)}_{q_{1}q_{2}}=\frac{1}{2}
  \left( \bar{q_1}_L \gamma^\mu {q_1}_L \right)
  \left( \bar{q_2}_L \gamma^\mu {q_2}_L \right) ,
  \nonumber\\
&&{\cal O}^{(1)}_{q_{1}q_{2}}=\frac{1}{2}
  \left( \bar{q_1}_R \gamma^\mu {q_1}_R \right)
  \left( \bar{q_2}_R \gamma^\mu {q_2}_R \right) ,
  \nonumber\\
&&{\cal O}^{(1)}_{l f}=\frac{1}{2}
  \left( \bar{l_L} \gamma^\mu l_L \right)
  \left( \bar{f_L} \gamma^\mu f_L \right) ,
  \nonumber\\
&&{\cal O}_{l f}=\frac{1}{2}
  \left( \bar{l_R} \gamma^\mu l_R \right)
  \left( \bar{f_R} \gamma^\mu f_R\right).
\end{eqnarray}
Thus, the relations (\ref{4.24}) give rise to the following
dependences between the parameters $\alpha_i$:
\begin{eqnarray}\label{5.27}
\alpha^{(3)}_\phi &=&-2\alpha_{\partial\phi}=
  \frac{2\alpha^{(1)2}_{\phi e}}{\alpha^{(1)}_{ee}},\quad
\alpha_{\phi f}= \alpha^{(1)}_{\phi f}+
  \frac{2t_f \alpha^{(1)2}_{\phi e}}{\alpha^{(1)}_{ee}},\nonumber\\
\alpha^{(1)}_{q_1 q_2}&=& \alpha^{(1,1)}_{q_1 q_2}+
  \frac{4t_{q_1}t_{q_2}}{1+\delta_{q_1 q_2}}
  \left( \frac{\alpha^{(1)}_{\phi q_2}}{t_{q_2}}
   +\frac{\alpha^{(1)}_{\phi q_1}}{t_{q_1}}
   +2\frac{\alpha^{(1)2}_{\phi e}}{\alpha^{(1)}_{ee}}\right),\nonumber\\
\alpha_{l f}&=& \alpha^{(1)}_{l f}+
  \frac{4t_f}{1+\delta_{l f}}
  \left( \alpha^{(1)}_{\phi l}
   -\frac{\alpha^{(1)}_{\phi f}}{t_f}
   -2\frac{\alpha^{(1)2}_{\phi e}}{\alpha^{(1)}_{ee}}\right) .
\end{eqnarray}
Hence it follows that the arbitrary constants are to be the ones
parameterizing the processes
$\bar{f}_{1L}f_{1L}\to\bar{f}_{2L}f_{2L}$ and $\bar{e}_L
e_L\to\phi\phi$.

The relations (\ref{5.27}) hold if the only additional heavy
particle extending the SM is the Abelian $Z^\prime$ boson. Other
models may lead to other type of correlations between the
parameters $\alpha_i$. The advantage of the expressions
(\ref{5.27}) is their independence of the mechanism of arising of
the Abelian $Z^\prime$ in a specific theory describing physics at
energy scale $E\ge\Lambda$.

%%%%%%%%%%%%%%%%%%%%%%%%%%%%%%%%%%%%%%%%%%%%%%%%%%%%%%%%%%%%%%%%%%%%%
%
%  Discussion
%
%%%%%%%%%%%%%%%%%%%%%%%%%%%%%%%%%%%%%%%%%%%%%%%%%%%%%%%%%%%%%%%%%%%%%
\section{Discussion}\label{sec:discuss}

Let us discuss the main points of our analysis and further
perspectives. Usually, the EL (\ref{intr:1}) describing at low
energies physics beyond the SM contains a set of the operators
with parameters $\alpha_i$ assumed to be arbitrary numbers which
must be fixed in experiments. However, as we have demonstrated
above considering the additional $Z^\prime$ boson as extension of
the SM, the renormalizability of the underlying theory may result
in some correlations between $\alpha_i$ which have to be of
interest in searching for new physics effects. Most important is
that the derived relations (\ref{4.24}), being a consequence of
renormalizability formulated in the framework of scattering in the
external field, are independent of the specific underlying (GUT)
model.

The latter statement requires to be discussed in more detail. The
Lagrangian (\ref{model:1}) respects $\tilde{\rm U}(1)$ gauge
invariance but includes only the interactions of renormalizable
type. The terms of the form $\sim(\partial_\mu Z^\prime_\nu
-\partial_\nu Z^\prime_\mu)\bar{f}\sigma_{\mu\nu}f$ are omitted
because they, being generated at GUT (or some intermediate
$\Lambda^{\rm GUT}>\Lambda^\prime>\Lambda$) mass scale, are
suppressed by the factors $1/\Lambda^{\rm GUT},1/\Lambda^\prime$.
Hence, one can conclude that at low energies, $E\ll\Lambda$, the
renormalizable interactions are to be dominant. Therefore, the
only model dependent information in Eq. (\ref{model:1}) is the
specific values of the parameters $\tilde{Y}^L_f$, $\tilde{Y}^R_f$
and $\tilde{Y}_\phi$. The relations (\ref{4.24}) hold for
arbitrary values of them, as it was demonstrated in Sec.
\ref{sec:RG4ferm}. As a consequence, the relations (\ref{5.27})
also hold for arbitrary $\tilde{Y}^L_f$, $\tilde{Y}^R_f$,
$\tilde{Y}_\phi$ and are independent of them explicitly.

Here, it will be useful to stress once again the main features of
our approach as compared to the ones, where the $\alpha_i$
parameters are assumed to be arbitrary numbers. In the latter
case, the origin of the composed operators is completely hidden.
In the former one, we first came to the one step back in the
analysis and have taken into account the most important
consequences of the renormalizability: 1) at low energies, the
interactions of renormalizable type are dominant whereas the
nonrenormalizable interaction terms are suppressed by the
additional degree of $\Lambda^{-1}$; 2) for any renormalizable
interactions at low energies one can interrelate scattering
processes due to the renormalizability of scattering amplitudes in
the external field of heavy particle.

Now, let us consider a more general parameterization in Eq.
(\ref{model:1}). In principle, one could employ the most general
form of the generator
$\tilde{Y}_\phi\to\mbox{diag}(\tilde{Y}_{\phi,1},\tilde{Y}_{\phi,2})$
which could be just the part of the appropriate generator of the
underlying theory. Then, the RG relations can be used to obtain
all possible solutions for the numbers $\tilde{Y}_{\phi,i}$,
$\tilde{Y}^L_f$, and $\tilde{Y}^R_f$. As one can check, the two
sets of correlations are allowed. The first one,
\begin{equation}\label{dis:1}
  \tilde{Y}^R_f=0,\quad
  \tilde{Y}_{\phi,1}+\tilde{Y}_{\phi,2}=0,\quad
  \tilde{Y}^L_{f^\prime}+\tilde{Y}^L_f=0,
\end{equation}
describes the vector boson analogous to the SM field $A^3_\mu$.
Such a field may appear as a component of the non-Abelian gauge
field. The second solution leads to the relations derived in this
paper:
\begin{equation}\label{dis:2}
\tilde{Y}_{\phi,1}=\tilde{Y}_{\phi,2},\quad
\tilde{Y}^L_{f^\prime}=\tilde{Y}^L_f,\quad
\tilde{Y}^R_f=\tilde{Y}^L_f +t_f \tilde{Y}_\phi.
\end{equation}
They correspond to the Abelian vector boson considered. Thus, the
proposed method to relate the parameters $\alpha_i$ could be
applied to interactions of light SM particles with other heavy
particles (leptoquarks, for example) treated as the external
fields.

In Ref. \cite{sirlin} a general renormalization framework for the
${\rm SU}(2)_L\times{\rm U}(1)_Y\times\tilde{\rm U}(1)$ theory was
presented, and, in particular, the set of relevant counterterms
has been derived. Owing to $\tilde{\rm U}(1)$ symmetry the same
counterterms were postulated for different $Z^\prime$ couplings to
fermions. However, as it was shown above, this fact requires the
specific relations between the hypercharges $\tilde{Y}^L_f$,
$\tilde{Y}^R_f$, $\tilde{Y}_\phi$. Naturally, the relations hold
automatically for a specific renormalizable higher-energy theory.
However, if the hypercharges are treated as unknown parameters
these relations are to be taken into account in order to respect
the gauge symmetry.

In fact, the obtained relations (\ref{4.24}) demonstrate that the
fermion and the scalar sectors of new physics are strongly
connected. The couplings of the $Z^\prime$ boson to the SM
axial-vector currents are completely determined by its interaction
with the scalar fields, if a single light scalar doublet is
assumed. When the $Z^\prime$ coupling to the scalar doublet is
switched off ($\tilde{Y}_\phi=0$), the $Z^\prime$ boson interacts
at the tree level with the SM vector currents, only
($\tilde{Y}^R_f-\tilde{Y}^L_f=0$). But even in this case the
couplings to the axial currents are to be produced by loops being
suppressed by the additional small factor $\sim g^2/16\pi^2$.

As it follows from Eq. (\ref{4.24}), the RG relations respect the
$\tilde{\rm U}(1)$ symmetry of the Yukawa terms responsible for
generating the masses of the SM fermions. Therefore, assuming the
basis low-energy theory to be, for example, the two-Higgs-doublet
SM, one could expect that the relations identical to Eq.
(\ref{4.24}) can be derived for the Abelian $Z^\prime$ couplings
to each of the scalar doublets. However, in general, the RG
relations are not determined by the specific gauge invariance of
the Lagrangian. For instance, consider the non-Abelian $Z^\prime$
couplings described by the solution (\ref{dis:1}). As one can
check, the relations (\ref{dis:1}) require no specific gauge
invariance of the Yukawa terms.

As it was mentioned in Sec. \ref{sec:model}, the $Z^\prime$
interaction with the scalar doublet is completely determined by
the observable $m^2_Z-m^2_W/\cos^2\theta_W$ (see Eq.
(\ref{model:5a})). Therefore, one is able to predict the coupling
of the $Z^\prime$ to the SM axial-vector currents. When $Z^\prime$
boson does not interact with the scalar doublet, the physical mass
of $Z$ boson is to be identical to its SM value. It is worth
noticing that the experimental value of $\cos^2\theta_W$ in Eq.
(\ref{model:5a}) must be properly chosen in order to satisfy the
definition $\tan\theta_W=g^\prime/g$. The $Z^\prime$ existence has
not to affect the quantity $\cos^2 \theta_W$ in order $\sim
m^2_W/\Lambda^2$. So, the determination of the Weinberg angle in
terms of the QED coupling constant $\alpha$ and the Fermi constant
$G$ measured from the muon decay is preferable, since the tree
level values of these parameters $\alpha =g^2\sin^2\theta_W/4\pi$,
$G=\sqrt{2}g^2/8m^2_W$ include the contributions of $W$ bosons and
the QED sector, only.

The proposed method can be applied to find the relations of the
derived type for different effective four-fermion operators
generated by the $Z^\prime$ (for instance, discussed in Refs.
\cite{hagiwara}). In general, it gives possibility to search for
other heavy virtual states that is the problem for future.

The authors are grateful to S. Dittmaier, D. Kazakov, A. Pankov,
N. Paver, D. Schildknecht, and O. Teryaev for discussions.

%%%%%%%%%%%%%%%%%%%%%%%%%%%%%%%%%%%%%%%%%%%%%%%%%%%%%%%%%%%%%%%%%%%%%
%
%  The Bibliography
%
%%%%%%%%%%%%%%%%%%%%%%%%%%%%%%%%%%%%%%%%%%%%%%%%%%%%%%%%%%%%%%%%%%%%%

%%%%%%%%%%%%%%%%%%%%%%%%%%%%%%%%%%%%%%%%%%%%%%%%%%%%%%%%%%%%%%%%%%%%%
%
%  Figures
%
%%%%%%%%%%%%%%%%%%%%%%%%%%%%%%%%%%%%%%%%%%%%%%%%%%%%%%%%%%%%%%%%%%%%%
\newpage
\end{document}